\documentclass[
    ,final            
  ]
  {aipproc}

\layoutstyle{6x9}

\begin{document}

\title{Octetonium at the LHC}


\author{Chul Kim}{
  address={Department of Physics, Duke University, Durham, NC 27708, USA}
}
\author{Thomas Mehen}{
  address={Department of Physics, Duke University, Durham, NC 27708, USA}
}

\begin{abstract}
Several models of new physics, such
 as grand unified theories, Pati-Salam models, chiral color models, etc., predict
the existence of 
an $SU(2)_L$ doublet of color-octet scalars (COS).  In the Manohar-Wise model,  the Yukawa couplings of the COS
are assumed to be  consistent with Minimal Flavor Violation  ensuring constraints from flavor physics are 
satisfied even for relatively light scalars. In this simple model  we consider the production 
of color singlet bound states of COS that we call octetonium. Octetonium  are mainly produced 
via gluon-gluon fusion and have significant production cross sections at the LHC. 
They can decay to pairs of gluons or electroweak gauge bosons. If the masses of the octetonia are 
1 TeV or less, these states will be visible as resonances in $\gamma \gamma, W^+W^-, ZZ$, and $ \gamma Z$. 
\end{abstract}

\maketitle


In the Standard Model (SM) electroweak symmetry breaking is accomplished by the introduction of a single 
color-singlet, $SU(2)_L$ scalar doublet. Over the years many extensions of the SM with richer
scalar sectors have been proposed and testing these models is an important goal of LHC physics. 
One possibility is a scalar sector containing $SU(2)_L$ doublets that are octets rather than singlets under $SU(3)_c$. 
Many models of new physics contain such particles including grand unified theories, Pati-Salam models, chiral color models, etc.~\cite{newphysics}.

An important constraint on models of new physics is the absence of flavor changing 
neutral currents (FCNC) at tree level. Generic theories with TeV-scale particles 
that can  lead to  FCNC's are ruled out. One way to ensure that such unwanted effects are not
present in new physics models is to impose Minimal Favor Violation (MFV)~\cite{MFV}.  In MFV one demands that 
all violation of the flavor symmetry of the SM be proportional to the SM Yukawa couplings.
Manohar and Wise~\cite{Manohar:2006ga} asked what is the most general scalar sector that is consistent with MFV and found the answer is quite restrictive: the only allowed scalars are those with quantum numbers $\bf (1,2)_{1/2}$ and $\bf(8,2)_{1/2}$,
i.e. the SM Higgs and color-octet scalars (COS). The simplest model containing COS
has the following Yukawa couplings to quarks~\cite{Manohar:2006ga}
\begin{eqnarray}\label{Yukawa}
{\cal L}_Y = - \eta_U \, g_{ij}^U \bar{u}_{R i} T^A Q_{L j} S^A - \eta_D \, g_{ij}^D \bar{d}_{R i} T^A Q_{L j} S^{A\,\dagger} + \mathrm{h.c.}, 
\end{eqnarray}
where 
\begin{equation} 
S^A =\left( \begin{array}{c} S^{A+} \\ S^{A0}\end{array}\right)
\end{equation} 
are the complex color-octet scalar fields, $Q_{L j}$ is $SU(2)_L$ doublet of quark fields, $u(d)_R$ is an up(down)-type right-handed quark fields. $g_{ij}^{U,D}$ are SM Yukawa coupling matrix elements and $\eta_{U,D}$ are unconstrained complex parameters. Three mass eigenstates of the color-octet scalar fields are $S^{A\pm}$, $S^{A0}_R = \sqrt{2} \, {\rm Re}\,S^{A0}$, and $S^{A0}_I = \sqrt{2}\, {\rm Im}\,S^{A0}$, whose masses at tree level are given by 
\begin{eqnarray}\label{masses}
m^2_{S^\pm} &=& m_S^2+\lambda_1 \frac{v^2}{4},  \\
m^2_{S_R^0} &=& m_S^2+(\lambda_1 +\lambda_2 +2 \lambda_3)\frac{v^2}{4}\, ,\\
m^2_{S_I^0} &=& m_S^2+(\lambda_1 +\lambda_2 -2 \lambda_3)\frac{v^2}{4} \, ,
\end{eqnarray} 
where $v$ is the vacuum expectation value of the Higgs field, and $m_S$, $\lambda_1$, $\lambda_2$, and $\lambda_3$ are parameters in the scalar potential~\cite{Manohar:2006ga}. 

The present experimental constraints on the masses are rather weak. From the CDF search $H\to b\bar{b}$ plus associated additional $b$ jets~\cite{CDF8954}, Ref.~\cite{Gerbush:2007fe} estimated that the  lower mass bounds for the COS is $\sim 200$~GeV~ when both the couplings $\eta_{U}$ and $\eta_{D}$ are of order unity. But masses lower than 200~GeV are allowed if $\eta_D$ is much less than $\eta_U$~\cite{Burgess:2009wm}. The production cross section for 
color-octet particles is much larger at the LHC than at the Tevatron because of the larger energy and 
higher luminosity for gluon-gluon fusion initiated processes. For color-octet masses in the range 
200 - 500 GeV the pair production cross section is of order 10-100 pb ~\cite{Manohar:2006ga}.

In this proceeding we consider the possibility of forming color singlet bound states of COS which we call octetonium. 
Formation of bound states is possible if the Yukawa couplings in Eq.~(\ref{Yukawa}) are $O(1)$ or smaller.
The octetonium production cross section are only an order of magntude below the pair production cross sections. 
The dominant decay mode for these states is to two gluon jets. They can also decay to pairs 
of electroweak gauge bosons, such as $W^+ W^-,\gamma \gamma, \gamma Z^0$ and $Z^0 Z^0$. We will see below that if the COS mass is less than 500~GeV, the cross section for $\gamma\gamma$ is large enough that the octetonia 
would appear as resonances above the SM background, and we expect this to hold for other final states as well.

Since the COS masses are much larger than $\Lambda_{\rm QCD}$, we expect that they will be Coulombic in 
nature and perturbation theory will be a good guide to some of their properties. Assuming this, the typical 
velocity of the COS within the octetonia is given by $v \sim N_c \alpha_s(m_S v)$, where $N_c$ is the number of colors.
For $m_S = 200-1200$~GeV, we find $v=0.37-0.30$ and $\alpha_S (m_S v) = 0.12-0.10$. The coupling and velocity are similar to what is found in bottomonium. In order to produce bound states, the formation time must
be shorter  than the lifetime of the constituents. Under the assumption that the main decay channels are $S^{A+} \to t\bar{b}$ and $S_{R,I}^{A0} \to t\bar{t}$ via the Yukawa couplings, the ratio of the formation time to the life time for $S^{A\pm}$ is\begin{equation}\label{formlife}
 \frac{\tau_{\rm form}}{\tau_{\rm life}} = \frac{2\Gamma[S^{A\pm}]}{B.E.} \sim 0.08\, |\eta_U|^2-0.6 \,|\eta_U|^2 \, ,
\end{equation} 
for $m_{S^\pm}$ in the range 200-500~GeV. Here $B.E.$ is the binding energy which is estimated to
be $m_S N_c \alpha_s(m_cv)^2/4$. Similar results hold for the time ratio of the neutral color-octet. So when $|\eta_U|< 1$, we see that the formation of the bound state is possible. 

\begin{figure}[!t]
\includegraphics[width=16cm]{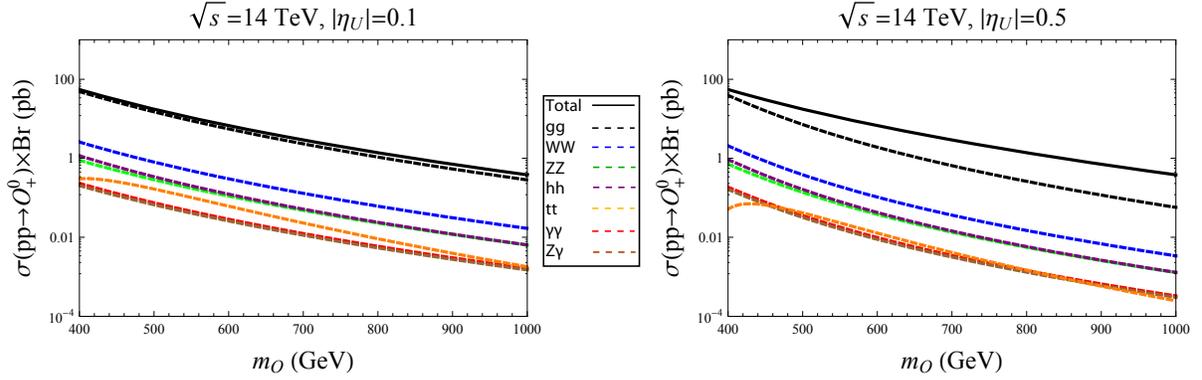} 
\caption{LHC cross sections for $pp\to O_+^0  \to X$, where $X$ is a two-body final state.\vspace{-3ex}}
\label{CrossSections}
\end{figure}

In this proceeding we will consider only three octetonia such as $O^0_+, O_R^0$, and $O_I^0$ which are the color-singlet bound states of $S^{A+}S^{A-}$, $S_R^{A0}S_R^{A0}$, and $S_I^{A0}S_I^{A0}$ respectively. The octetonium wavefunction
is
\begin{equation}\label{wfn}
|O^0_+(P)\rangle = \frac{\delta^{AB}}{\sqrt{N_c^2-1}}\sqrt{\frac{2}{m_O}}\int \frac{d^3 k}{(2\pi)^3} \, \tilde \psi(k)\, |S^{A+}(P/2+k) S^{B-}(P/2-k)\rangle ,
\end{equation} 
for $O^0_+$, where $P^\mu$ is the momentum of  $O_+^0$,
$k^\mu$ is the relative momentum of the $S^A$ in the bound state, and $\tilde \psi(k)$ is the momentum space wave function. For $O_{R,I}^0$, the same definition can be applied, but divided by $\sqrt{2}$. 

At parton level, the scattering cross section for the bound state is 
\begin{equation} 
\hat{\sigma}[g g \to O^0_+] = 2\hat{\sigma}[g g \to O^0_{\rm R}] =\frac{9\,\pi^3 \alpha_s^2(2 m_S)}{2 m_S \hat{s}}  |\psi(0)|^2 \, \delta(\hat{s} - m_O^2) \,  
\end{equation} 
where $\hat{s}$ is the partonic center of mass energy squared. For the wave function at the origin  squared 
we use the Coulombic approximation 
\begin{equation}
|\psi(0)|^2 = \frac{N_c^3 \alpha_s^3(m_S v) m_S^3}{8\pi}.
\end{equation} 

If the Yukawa coupling $\eta_U$ is smaller than one so that the $S^{A}$ are sufficiently long lived, 
the dominant decay channel  for the octetonia is via  annihilation of COS to two gluons. However, it is doubtful
that they could be discovered as a resonance in the two jet cross section since the SM background 
is enormous. More promising would be to search in rare decays to final states with two electroweak gauge bosons such as 
$\gamma\gamma$, $W^+W^-$, $\gamma Z^0$, and $Z^0Z^0$. For these final states backgrounds are much smaller 
than for two jets. 
In Fig.~\ref{CrossSections}, we show the scattering cross sections for $gg\to O_+^{0} \to X$ where $X$ are several two body final states. The branching fractions and the analytic expressions for the decay rates can be found in Ref.~\cite{Kim:2008bx}.

\begin{table}[!t]
  \begin{tabular}{ccccc}
\hline
   &  $400~{\rm GeV}$ \hspace{0.1 in} &   $600~{\rm GeV}$ 
   &  $800~{\rm GeV}$ \hspace{0.1 in}  &   $1000~{\rm GeV}$   \\
\hline
\hspace{0.1 in}$\sigma_{\rm res}[pp \to O_+^0 \to \gamma \gamma](\mathrm{fb}), \,|\eta_U|=0.1$\hspace{0.1 in} & 187  & 24  &  5.1  & 1.4 \\
\hline
\hspace{0.1 in}$\sigma_{\rm res}[pp \to O_+^0 \to \gamma \gamma](\mathrm{fb}), \, |\eta_U|=0.5$\hspace{0.1 in} & 149  & 7.7  &  1.1  & 0.23 \\
\hline
 $\sigma_{\rm{SM}}[pp \to \gamma \gamma] (\mathrm{fb})$ & 12.0 & 2.2 &0.64  & 0.23  \\
\hline 
\end{tabular}
\caption{\baselineskip 3.0ex
$\sigma_{\rm res}[pp \to O_+^0 \to \gamma \gamma]$ is the resonant cross section for photons with $|\eta|<2.4$ and $m_{\gamma \gamma}$
within $\pm 3$ GeV of the resonance mass for different values 
of $m_{O^0_+}$, $\sigma_{\rm SM}[pp \to \gamma \gamma]$ is the SM contribution in the same region.
}
\label{table}
\end{table}

To check if the signal for the bound states will be visible above SM backgrounds we calculate the differential scattering cross section for $pp\to O_+^0 \to \gamma\gamma$ process with respect to $m_{\gamma\gamma}$. Then we integrate this in the range ($m_{\gamma\gamma}- \Delta/2,~m_{\gamma\gamma}+ \Delta/2$) employing the Breit-Wigner profile. Here we chose $\Delta = 6~\rm{GeV}$ which is  the expected energy resolution below $m_{\gamma\gamma} = 1~\rm{TeV}$ in the ATLAS experiment \cite{:2008zzm}. This is compared with the SM contribution to this process
in Table~\ref{table}. The signal is dominant over the SM backgrounds for octetonium in the mass range 400 - 1000 GeV. 

To summarize, we have argued that  the COS can form strongly bound color-singlet states called octetonium
provided the Yukawa couplings of the COS to SM quarks is $O(1)$ or smaller. This condition is required
so that the COS do not decay to heavy quarks before they have time to form a resonance via their strong Coulomb
attraction. We have computed production cross sections for octetonia at the LHC which turned out to be quite large
as well as two body-decay rates to SM particles. The decay to two gluons dominates over 
others, but rare decays to SM electroweak bosons are also significant. In particular we showed that the process
$pp \to O_+^0 \to \gamma \gamma$ will be visible as a resonance above the SM background when the octetonia 
masses are in the range 400-100 GeV. Though we have not done explicit calculations, 
we expect similar results for  $W^+W^-$, $\gamma Z^0$, and $Z^0Z^0$. 
The higher order QCD corrections including soft gluon 
resummation for the precise prediction of total and differential scattering cross sections will be available in the near future \cite{IKM}.


\begin{theacknowledgments}
\vspace{-0.3cm}
This work was supported in part by the U.S. Department of Energy under 
grant numbers DE-FG02-05ER41368 and DE-FG02-05ER41376. We thank Ahmad Idilbi for useful discussions. 
\end{theacknowledgments}


\bibliographystyle{aipproc}   


\end{document}